\documentclass[11pt]{article}
\usepackage[margin=1in]{geometry}

\usepackage{amstext,amsmath,amssymb}
\usepackage{sectsty}
\usepackage{graphicx}

\def\barr{\hbox{I\hskip -.5ex R}}

\newcommand\Pen{\mbox{\text{Pe}}\,}  
\renewcommand{\Re}{\operatorname{Re}}

\title{Mass distribution and skewness for passive scalar transport in pipes with polygonal and smooth cross-sections}
\author{Manuchehr Aminian, Roberto Camassa, Richard M. McLaughlin}
\date{}

\begin{document}

\maketitle

\begin{abstract}

We extend our previous results characterizing the loading properties of a diffusing passive scalar advected by a laminar shear flow in ducts and channels to more general cross-sectional shapes, including regular polygons and smoothed corner ducts originating from deformations of ellipses.  For the case of the triangle, short time skewness is calculated exactly to be positive, while long-time asymptotics shows it to be negative.  Monte-Carlo simulations confirm these predictions, and document the time scale for sign change.  Interestingly, the equilateral triangle is the only regular polygon with this property, all other polygons possess positive skewness at all times, although this cannot cannot be proved on finite times due to the lack of closed form flow solutions for such geometries.  Alternatively, closed form flow solutions can be constructed for smooth deformations of ellipses, and illustrate how both non-zero short time skewness and the possibility of multiple sign switching in time is  unrelated to domain corners.  Exact conditions relating the median and the skewness to the mean are developed which guarantee when the sign for the skewness implies front (back) loading properties of the evolving tracer distribution along the pipe.  Short and long time asymptotics confirm this condition, and Monte-Carlo simulations verify this at all times.
\end{abstract}

\section{Introduction}

Recently, we established the surprising role which the cross-sectional tube shape plays in establishing the symmetry properties of a diffusing passive scalar advected by a laminar shear flow \cite{Science2016,PRLSkew15}.  Specifically, using a combination of analysis, simulations, and physical experiments, it was shown that the aspect ratio alone may be used to control the upstream-downstream solute distribution whereby skinny tubes yield distributions which are front loaded, having more mass arriving at the target before the mean arrives on long, diffusive time scales.  This property is reversed for fat domains.  This result was obtained for tubes of either elliptical or rectangular cross-section, and the critical aspect ratio, approximately 2-1, was computed {\it exactly} for elliptical cross-sections.  On very short timescales relative to the diffusion timescale, it was established that elliptical domains preserve symmetries, while rectangular domains break symmetry instantly, with a similar sorting as occurring at long times, with a numerically computed critical ratio, similarly near 2-1.  On intermediate timescales, high fidelity Monte-Carlo simulations showed quite complex dynamics, exhibiting loading properties which shift the loading front versus back loading properties, sometimes multiple times depending upon the aspect ratio.

The diffusive timescale, $L^2/\kappa$, is roughly the time it takes a particle to diffuse across the cross-sectional domain.  For practical applications in micro-fluidics, this timescale may occur quickly, on the order of seconds in experiments which may last minutes, which makes the long-time analysis quite relevant.  However, in many applications even in microfluidics, the pre-mixing phases occurring before the diffusion timescale is reached may be important, particularly in designing complex chemical reactions which need to occur well before the target is reached.  Further, while elliptical and rectangular ducts are likely the ideal in many micro-fluidic applications, advances in lithography and 3D printing have opened new avenues for exploration.

With this in mind, here we examine the symmetry properties induced by cross-sectional shapes beyond those of rectangular ducts, and elliptical tubes.  Specifically, we consider first regular polygons.  It may be shown that the only domain for which the fluid flow may be expressed as a polynomial in the cross-sectional coordinates is the equilateral triangle.  All other polygons require infinite series representations, or other numerical constructions of the fluid flow.  In turn, the analysis to compute the long time skewness asymptotics was generally developed in our prior work \cite{Science2016}, and relies upon the inversion of a family of nested Poisson problems, similarly requiring numerical inversion in polygonal domains.  We apply this analysis, using finite element solvers, and establish remarkably that the only polygonal domain yielding negative long time solute negative skewness is the triangle.  We additionally develop a generalized Monte-Carlo solver capable of handling a wide class of complex domains including polygons which can explore the intermediate time behavior, and crucially, can verify that indeed the sign of the skewness indeed indicates a proper ordering of the median and the mean.  This is important because there are known examples demonstrating that in general there is no direct correlation between skewness, median, and mean.

We develop an exact mathematical criterion which guarantees under general situations that the sign of the skewness indeed guarantees the front versus backloading properties of the solute distribution.  This criterion is shown to be satisfied generically at both short and long times.  Interestingly, this criterion is seen to not be satisfied on intermediate timescales in our Monte-Carlo simulations, but the correlation between skewness, median, and mean is nonetheless directly observed to be maintained.  Lastly, following our prior work \cite{PRLSkew15} we explore a family of deformations of elliptical domains which induce exact closed form flow solutions.   These solutions are built using complex analysis in which a harmonic function is added to the complexified solutions governing the elliptical domains.  The zero level set of the sum provides a new domain.  This new domain is seen to yield a wide range of smooth shapes offering effectively smoothed rectangles, and even non-convex cross-sections.  We apply our analysis and Monte-Carlo simulations to these new domains and demonstrate quite robustly that the 2-1 aspect ratio separating front from back loaded solute distributions extends beyond the elliptical and rectangular domains previously considered.

\section{Passive Transport}

The concentration evolution of a passive scalar in a prescribed laminar shear flow is given by the solution to the follow PDE:
\begin{equation}
\frac{\partial T}{\partial t}+ \Pen u(y,z) \frac{\partial T}{\partial x}=\Delta T
\end{equation}
where $T$ is the scalar concentration, and $u(y,z)$ is the pressure driven shear flow given by the solution to the Poisson equation:
\begin{equation}
\Delta u= -2
\end{equation}
where the nondimensionalization is the same as given in \cite{Science2016}.  The parameter, $\Pen$, is the Peclet number, which gives the relative importance of fluid advection to molecular diffusion.  In this nondimensionalization, time is measured in units of the the diffusion timescale defined above.  The boundary conditions for the field $u$ are vanishing and for the field $T$ are vanishing Neumann on the curve defining the cross-section.

Longitudinal spatial moments of the concentration distribution characterize its width, and symmetry properties.  These may found carefully described by the Aris moment equations in \cite{Aris56,Science2016,PRLSkew15}.  The main statistic we focus on below is the skewness, which is centered, normalized third moment.  $\langle(X-\langle X\rangle)^3\rangle /(\langle X-\langle X\rangle)^2\rangle^{3/2}$, where $\langle \cdot \rangle= \int_{R^1} (\cdot) T(x,y,z) dx$, unless otherwise defined.

\section{Triangular cross section}
The equilateral triangle is the only regular polygon
in which the flow has a closed form expression, hence
we start with it. Using the polynomial flow solution,
we compute exact short asymptotics, while the long time
asymptotic coefficients are computed using a finite element package.
Then we compare these results to Monte Carlo simulations
and see excellent agreement.

\subsection{Exact solution for shear flow in triangular cross section}

In the case of a cross section which is an equilateral triangle,
there is an exact formula for the flow:
\begin{equation}
\label{eqn:triangle_flow}
u(y,z) = \frac{1}{12a}(a+y)(2a+\sqrt{3}z-y)(2a-\sqrt{3}z-y)
\end{equation}
where $a$ is defined to be the distance from the centroid
(fixed at $(0,0)$) to the nearest boundaries, hence
each side is length $2a\sqrt{3}$. This choice of length scale
$a$ is chosen so that there is consistency between this geometry
and the previously studied infinite channel, rectangles, and
ellipses.  It is an elementary calculation to verify that this shear flows satisfies the Navier-Stokes equations with vanishing boundary conditions under a uniform pressure gradient, which reduce to a Poisson equation with constant driver.
Each term in the product \ref{eqn:triangle_flow}
plays the role of imposing Dirichlet
boundary conditions on the respective side of the triangle:
\begin{equation}
y=-a, \quad y=2a + \sqrt{3} z, \quad y=2a - \sqrt{3} z
\end{equation}
which verifies the Dirichlet boundary conditions are satisfied.
The scaling is chosen so that the Laplacian of the solution is $-2$. Nondimensional
scaling is equivalent to setting $a=1$ here.

\subsection{Monte Carlo for convex polygonal boundary}
\label{subsec:tri_mc}
Here we discuss some of the details for efficient
implementation of applying reflecting boundary
conditions in the case of a general convex polygon, for which the triangle is a special case.

It is assumed the flow $u(y,z)$ is already known (as
in the triangle), so that the two other steps,
detecting a boundary crossing and applying a reflection,
are all that remains.

For this case, the Monte Carlo code is roughly a
modification between previous implementations in the
infinite channel (since the flow is polynomial)
and rectangular cases (where there is a bounded
two-dimensional cross section), for details see \cite{Science2016,PRLSkew15,AminianThesis}.
However, it is not quite as trivial to apply boundary
conditions as in the channel or rectangle (where
only subtractions are needed).

Our approach is to
describe the triangle as an intersection of the three
half-planes $\ell_i(y,z)>0$, where each $\ell_i$ is
essentially one of the boundaries written above:
\begin{subequations}
\begin{align}
\ell_1(y,z) &= a + y, \\
\ell_2(y,z) &= 2a -y + \sqrt{3} z, \\
\ell_3(y,z) &= 2a -y -\sqrt{3} z.
\end{align}
\end{subequations}
or for a generic linear boundary,
\begin{equation}
\ell(y,z) = c_0 + c_1 y + c_2 z,
\end{equation}
with specified coefficients $c_0$, $c_1$, $c_2$.
Note that there is a freedom of
the overall sign, i.e., $\ell_1 = \pm (a+y)$ both
capture the boundary.
Our convention is to generally choose the sign so that
$\ell_i(y,z)$ is positive if $(y,z)$ is
in the domain. If we assume $(0,0)$ is in the domain,
then this is the requirement that $\ell_i(0,0)>0$;
for this case, $\ell_1 = +(a+y)$. Generally this means
choosing the sign so that the constant term is positive.

The fact that each piece of boundary is a line means the
normal direction $\underline{n} \propto \nabla \ell_i$
is given for free from the coefficients of $\ell_i$.
For instance,
if we had only the half-space advection diffusion
problem with boundary $\ell=0$, after seeing a
particle go from $(y_0,z_0) \rightarrow (y_1,z_1)$
with $\ell(y_1,z_1)<0$ (hence, exiting the domain),
the main steps are to, first,
calculate the time of intersection by solving
\begin{equation}
\ell\big(y_0 + s (y_1-y_0), z_0 + s (z_1 - z_0) \big)=0.
\end{equation}
This has an
exact solution for arbitrary $\ell(y,z) = c_0 + c_1 y + c_2 z$:
\begin{equation}
\label{eqn:triangle_ssol}
s = \frac{c_0 + c_1 y_0 + c_2 z_0}{c_1(y_0 - y_1) + c_2(z_0 - z_1)} =  \frac{\ell(y_0,z_0)}{c_1(y_0 - y_1) + c_2(z_0 - z_1)}.
\end{equation}
Next, find the normal direction at the intersection point.
Again, this has an exact solution, but it is also
position independent:
$\nabla \ell = ( c_1, c_2 )$.
Finally, apply the appropriate reflection about the tangent
plane, preserving the distance traveled,
$|\langle y_1 - y_0, z_1 - z_0 \rangle|$.
\begin{figure}
\centering
\label{fig:polygon_reflections}
\includegraphics[width=\linewidth]{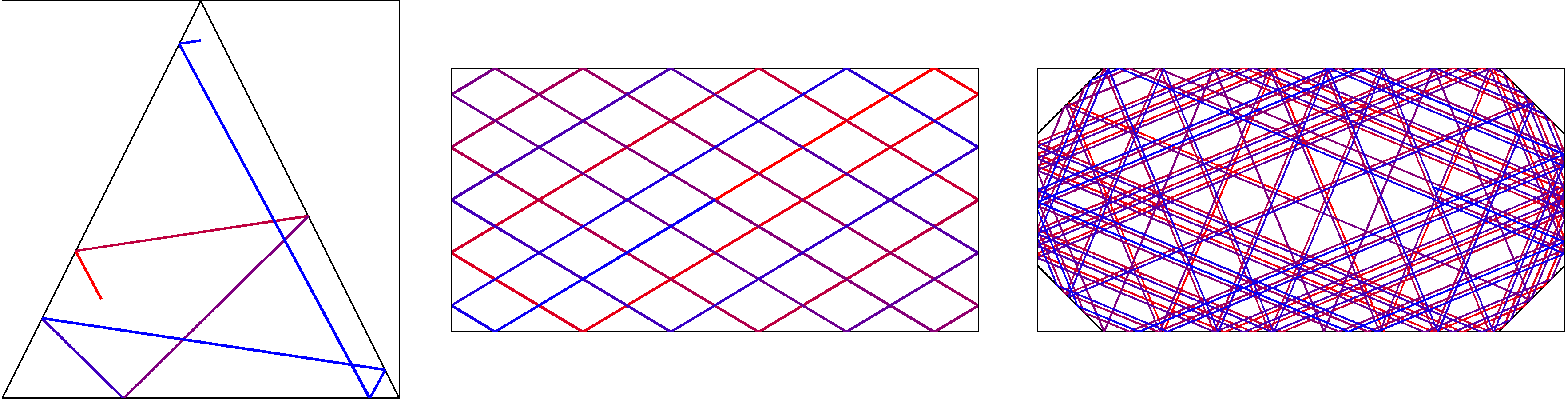}
\caption{
Demonstration of the reflection algorithm for
some convex polygons. In the case where an extremely long trajectory places the first reflection still outside the domain, then the reflection algorithm is applied
iteratively until
the final position $(y_1,z_1)$ returns to the domain.
The number of reflections is illustrated in
the changing color.
Left: equilateral triangle with eight reflections.
Center: Reflection in a rectangle $\lambda=1/2$
whose initial outward trajectory has a rational
slope. Right: demonstration
in a non-regular octagon of the same
aspect ratio.
}
\end{figure}

If multiple boundaries are crossed on a single
timestep, one needs to
check the formula (\ref{eqn:triangle_ssol}) for
all $\ell_i$, take the smallest positive value of
$s$ and the appropriate $\ell_j$, reflect about the line
$\ell_j=0$, and repeat this process until $\ell_i(y_1,z_1)>0$ for all $i$
(a trajectory is in the domain if and only if $\ell_i(y,z) >0$
for all $i$).
An implementation of this is shown in figure
\ref{fig:polygon_reflections}, with a single large trajectory
for the triangle (left) appropriately handled as
many reflections, starting from red and ending with blue.

To show how this can be generalized, the center and
left panels of figure \ref{fig:polygon_reflections}
implement the same algorithm in a rectangle and an irregular
octagon. The main algorithm is identical; one only needs
to specify additional planes $\ell_i$. The
trajectory is given a rational slope in the rectangle
to verify angles are preserved properly, while the
octagon is an implementation in a non-trivial domain, and
shows interesting ``bimodal" behavior due to the diagonal
boundaries.
Though not shown here, the total distance traveled
without boundaries, and with boundaries
after applying reflections (as the
sum of lengths of line segments),
is verified to be conserved.
Otherwise, the same time integrator described in our prior work is utilized to construct the Monte-Carlo sampling.  Unless otherwise stated, the maximum timestep employed is $0.0001$, and $50$ simulations each using $10^6$ random walkers uniformly distributed across the cross-section.

\subsection{Calculation of asymptotics}

\begin{figure}
\centering
\includegraphics[width=\linewidth]{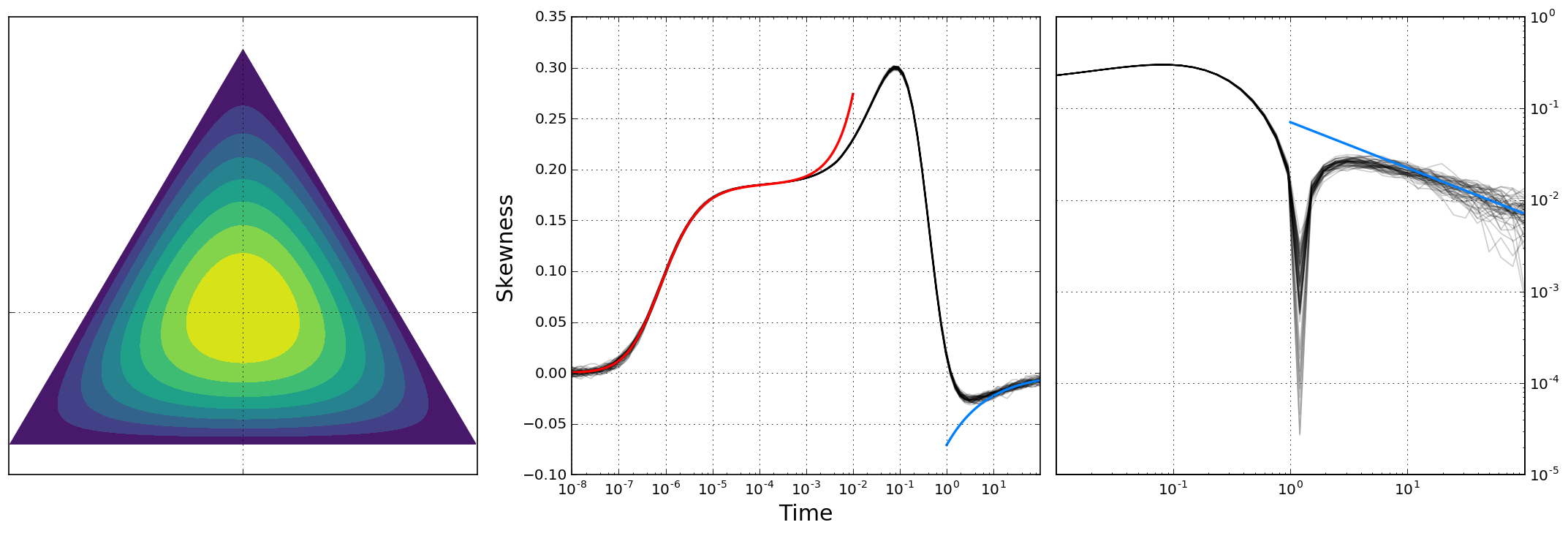}
\caption{
Results in the triangular geometry. Left:
schematic of the flow profile, with $y=0$ and $z=0$ lines.
Center: skewness in fifty simulations (black), each with $10^6$ random walkers,
and short time (red) and long time (blue) asymptotics,
with $\Pen=10^4$. Right: the same simulations and
long time asymptotics with log-scaled axes.
}
\label{fig:triangle_sim}
\end{figure}

Since the flow is a polynomial, the short time
asymptotics can be calculated exactly. After shifting to
the reference frame of the mean velocity so that
the average velocity $\langle u \rangle = 0$, the
geometric skewness \cite{PRLSkew15} describing the
skewness at short time induced by advection, is
calculated to be
\begin{equation}
\label{eqn:stskasymptotic}
\frac{\langle u^3 \rangle}{\langle u^2 \rangle^{3/2}} = \frac{1/19250}{(3/700)^{3/2}} = \sqrt{\frac{112}{3267}} \approx 0.185,
\end{equation}
which predicts positive short time skewness, i.e., back-loading.  The red curve shown in figure \ref{fig:triangle_sim} is given by the theory developed in \cite{PRLSkew15}, applied the exact flow in the triangle, which yields the temporal short time skewness evolution, via its moments:

\begin{equation}
\begin{array}{l}
\mathcal{M}_1 \sim 0, \: \mathcal{M}_2 \sim 2 \, \tau + \Pen^2 \langle u^2 \rangle \, \tau^2 - \frac{2}{3} \Pen^2 \, \langle \tilde{u} \rangle \, \tau^3, \\[2pt]
\mathcal{M}_3 \sim \Pen^3 \langle u^3 \rangle \, \tau^3 - \left( \langle \tilde{u}^2 \rangle - 2\langle \tilde{u} \rangle ^2 \right) \tau^4
\end{array}
\end{equation}
Here, $\tilde u$ denotes the flow vanishing at the boundary, while $ u = \tilde u-\langle \tilde u \rangle$, with here $\langle \cdot \rangle$ denoting cross-sectional average.

Incidentally, we note that this gives the leading order and first asymptotic correction at short times for the skewness evolution.  We previously established that the leading order coefficient vanishes for ellipses of all aspect ratios.  By providing an explicit generic formula for the first order correction, it further predicts the surprising fact that the short time skewness for {\em all} ellipses is positive, in strong contrast to the behavior in rectangular ducts for which skinny cases evolve negatively, whereas fat cases do the opposite.

The long time asymptotics are calculated using long time
asymptotics of the Aris equations \cite{Aris56}, a collection
of coupled, driven diffusion equations describing the flow-wise
tracer statistics.
The details are in the supplementary material of \cite{Science2016},
but we give a brief outline here.
The approach is to use the time-dependent,
long time asymptotics of the Aris equations, applying solvability
conditions to obtain leading-order behavior after
the last diffusive timescale. Generically, after the mean-zero
flow $u$ is known, the solution of two cell problems
are needed:
\begin{align}
-\Delta g_1(y,z) &= \, u(y,z), \quad \partial_{\underline{n}} g_1|_{\partial \Omega} = 0, \quad \langle g_1 \rangle = 0, \label{eqn:ltg1problem} \\[1em]
-\Delta g_2(y,z) &= 2 \big( ug_1 - \langle ug_1 \rangle \big), \quad \partial_{\underline{n}} g_2|_{\partial \Omega} = 0, \label{eqn:ltg2problem}
\end{align}
where $\langle \cdot \rangle$ denotes the cross-sectional average, and
$\underline{n}$ is the unit outward normal to the boundary.
From this the leading order behavior of skewness at long time is
\begin{equation}
\label{eqn:ltskasymptotic}
Sk(y,z,\tau) \sim Sk(\tau) = \frac
{3 \Pen^3 \langle u g_2 \rangle}
{\big( 2 + 2 \Pen^2 \left \langle u g_1 \right \rangle \big)^{3/2}} \; \tau^{-1/2},
\end{equation}
showing that the term $\langle u g_2 \rangle$ predicts the direction of skewness, while
the full ratio also indicates the relative magnitude preceding the final timescale.
Thus, to apply this theory, these cell problems need to be either solved analytically, or numerically.  We describe below the numerical methodology we use to solve these problems in arbitrary domains using finite element packages.  In our prior work, we established the exact solutions for flow in ellipses of arbitrary aspect ratio \cite{Science2016}.

Applying this theory in the present case using the methodology described below, we calculate the leading coefficient;
the result is \emph{negative}:
\begin{equation}
\frac{3 \langle u g_2 \rangle}{\left( 2 \langle u g_1 \rangle \right)^{3/2}} \approx -0.07076.
\end{equation}
The results are summarized in figure \ref{fig:triangle_sim}. The left panel is a
contour map of the flow velocity throughout the cross section. The center panel compares
the asymptotic predictions at short (red) and long (blue) times to a collection
of Monte Carlo simulations in the triangle. The right panel compares the same simulations
to the long-time prediction only. The axes are logarithmically scaled (using the absolute
value of the skewness) to show that both the rate of convergence $\tau^{-1/2}$
as well as the leading coefficient agree well with the simulations.

\section{Asymptotics in the regular polygons}

The results of the previous section showing
that the long time skewness in the triangle
is negative raises an interesting question as
to the other general polygons. We use a
finite element package to calculate the asymptotic
coefficients, and show that in fact the triangle
is the only one of the regular polygons for which
this occurs. This can be understood interpreting
the regular polygons as a sequence of geometries which
limits to the circle.
\begin{figure}
\centering
\includegraphics[width=\linewidth]{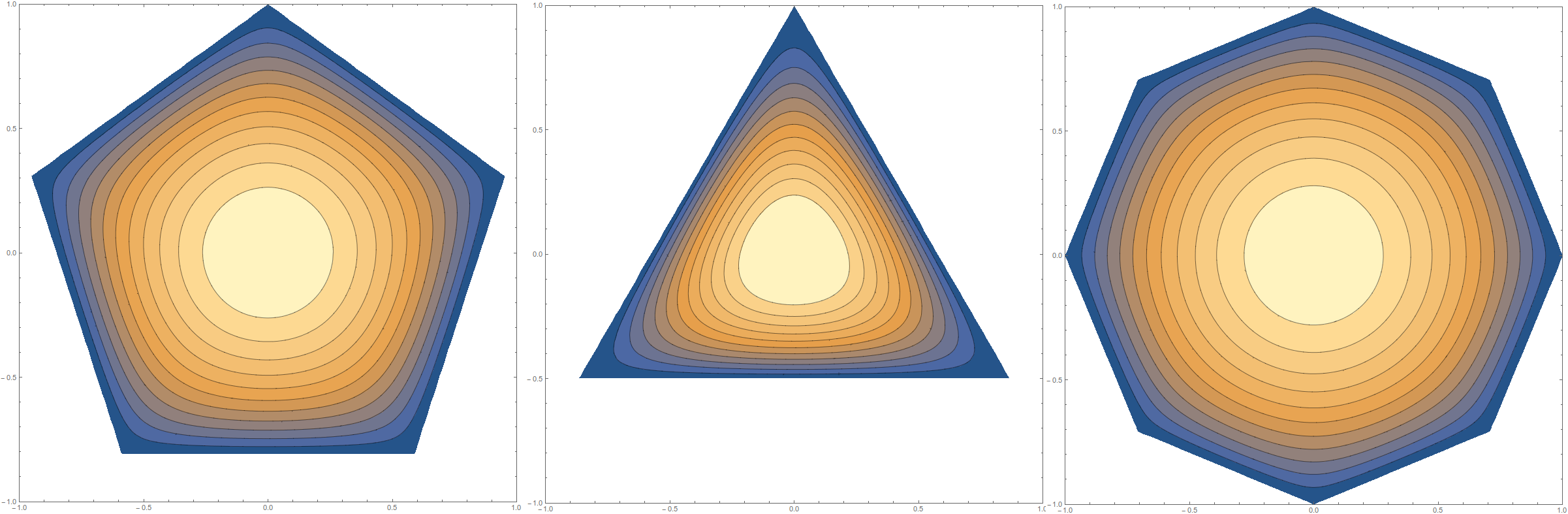}
\caption{
Visualization of the numerically computed flow
profile in the regular $n$-gons for $n=5, n=3,$ and $n=8$
respectively.
Curvature of the level sets is more influential
for $n$ small, but is almost immediately washed out
for $n$ large.
}
\label{fig:ngon_vis}
\end{figure}

To calculate the flow and coefficients of the
short and long time asymptotics in these domains,
we use the finite element package available
in Mathematica 10. The package allows
one to define domains explicitly or implicitly,
specify arbitrary boundary conditions and right-hand
sides, and control basic meshing and solver options.
We validated the finite element package against
our exact results for the coefficients in
the family of ellipses as reported in \cite{Science2016}.
We set the mesh parameters
\texttt{MaxCellMeasure $\to 10^{-3}$} and
\texttt{MeshOrder $\to 2$}; numerical integrations
were set with parameters \texttt{AccuracyGoal $\to 10$}
and \texttt{PrecisionGoal $\to 10$}.
The choices of parameters were
chosen after a convergence study to see
convergence of the asymptotic coefficients to
the number of digits reported.
The polygonal domains were defined using the
locations of their vertices on a circumscribed circle;
$(y_k,z_k) = (\cos(2\pi k/n),\sin(2\pi k/n))$,
$i=0,...,n-1$.
The flow solutions for a few polygons are visualized
in figure \ref{fig:triangle_sim}.

Using this tool, the flow and the analogous
cell problems to
(\ref{eqn:ltg1problem}) and (\ref{eqn:ltg2problem})
are calculated and the asymptotic coefficients
(\ref{eqn:stskasymptotic})
and (\ref{eqn:ltskasymptotic}) are calculated numerically.
The results are reported in table \ref{table:ngon_coeffs}.
Observe that the triangle is the only cross
section of this class to have a negative coefficient.
Aside from this, the coefficients are all positive and
monotonically approach the circular values ($n=\infty$).

Note that the triangular result differs from the
result calculated in section \ref{subsec:tri_mc}
as the value of $a$, the radius of the \emph{inscribed}
circle, is different when constructing
the $n$-gons as in this case, using the \emph{circumscribing}
circle. While the analagous value of $a$ will in fact
grow with increasing $n$ (limiting to $1$ as $n \to \infty)$,
so that care must be taken when comparing magnitudes of
statistics between these $n$-gons, the sign
of the coefficients does not depend on $a$.

The triangle being unique with negative skewness is
surprising. We give some intuition here. For
reference, we visualize the numerically computed flow
in figure \ref{fig:ngon_vis} for the triangle, pentagon,
and octagon.

As was discussed in our prior work \cite{PRLSkew15,Science2016}, the physical intuition regarding the interplay between advection and diffusion is quite subtle.  It may be shown that starting a point source initial condition (as opposed to the generally uniform initial data considered here) near the wall, produces a positive skewed distribution on short timescales (and negative for point sources located at the centerline). Alternatively, at long times, the memory of the initial data is essentially lost, and the sign of the skewness is a complicated balance between the interior and the wall.  One observation noteworthy in the present discussion is the difference in velocity contours between the different geometries:  The boundaries which have large (effective) curvature create larger regions of slow flow in the cross
section. For the cases $n=3$ and $n=5$,
we see higher curvature regions
of the level sets persist as one traces
a path from one of the corners towards the
center of the domain. However, for $n=8$, the
corners have little influence, and the level sets
are almost immediately circular. This is one way
of understanding the approach of the flow
profile (and thus the asymptotic coefficients)
to the values for the circular cross section.

\begin{table}
\centering
\begin{tabular}{| c | c | c | c | c | c | c | c | c | c | c | c |}
\hline
$n$ & 3 & 4 & 5 & 6 & 7 & 8 & 20 & 100 & $\infty$ \\
\hline
$\mathcal{S}^\mathcal{G}$ & 0.185 & 0.0813 & 0.042 & 0.024 & 0.0149 & 0.0097 & $3.97 \times 10^{-4}$ & $8.6 \times 10^{-7}$ & 0 \\
\hline
LT coeff & $-0.035$ & 0.139 & 0.194 & 0.215 & 0.225 & 0.231 & 0.243 & 0.245 & 0.245 \\
\hline
\end{tabular}
\caption{
Geometric skewness $\mathcal{S}^\mathcal{G}$ and the long time coefficient
$3\langle u g_2 \rangle /(2 \langle u g_1 \rangle)^{3/2}$ (third row)
calculated numerically for the regular $n$-gons.
The coefficients monotonically approach the
circle ($n = \infty$) value; only in the triangle case on
is negative asymptotic skewness seen.
}
\label{table:ngon_coeffs}
\end{table}

The intuition explaining negative skewness above
may raise a mathematical question about whether the
degree of continuity of the boundary (i.e., continuous but
not differentiable) plays a direct role. We will see
next that this is not the case: the aspect ratio near $2:1$ generically plays a fundamental role in determining the sign of skewness at long times.  At short times, however, the role of boundary in setting the skewness sign is more subtle:  ellipses of any aspect ratio have zero skewness (and become positive in the evolution), while ducts and their rounded counterparts acquire different skewness signs depending upon their aspect ratio, again with the critical ratio being near $2:1$. Of course the boundary acts to deform the velocity contours in non-trivial ways which could give rise to other interesting effects, as for instance in the case of the equilateral triangle.

\section{Racetrack cross sections}
The ``racetracks," as we have termed them, are
an extension of the elliptical cross sections.
The method as described in \cite{PRLSkew15,AminianThesis} to
construct them is to modify the ellipse flow solution,
taking the new cross section to be the zero level set
of this new flow.  In that prior work, amongst other observations, it was established that instantaneous symmetry breaking as characterized by the geometric skewness, not possible in elliptical domains, nonetheless is possible in domains with and without corners.

Each of these geometries has an
aspect ratio $\lambda$, as in the ellipses and rectangles,
as well as a shape parameter $s$, which to some degree
allows one to smoothly modify the domain from an
ellipse to rectangle-like domains.
Constructing a new family of cross sections
in this way is relatively straightforward;
however, some care must be taken to the domain of
$(\lambda,s)$ values to ensure
the new family of domains is physically relevant
and behaves as expected.

The goal in examining this family of cross sections
is to better understand the connection between the class
of rectangles and ellipses, which show distinctly different
behavior at short time, yet similar behavior at long time.
We have extended our Monte Carlo code to handle these geometries,
in addition to numerical calculation of
the asymptotic coefficients. Ultimately we will show that
the geometric skewness appears to depend continuously on the
shape parameter $s$, and that the skewness at long time has
only weak dependence on $s$. This could have important
implications for applications, as this suggests that the
details of the cross-sectional shape are not so important as
simply the aspect ratio itself,
if the experimental timescales one is interested in are
near the diffusive timescale.

\subsection{Derivation of the flow solution }
Usually, a Poisson problem is given with a specified
domain and boundary conditions in advance, and the
solution $u$ is found from this information.
The idea for this section is to modify
a known solution $u$ to the Poisson equation by
adding to it a harmonic function, which will still
satisfy $\Delta \tilde{u} = \textit{const.}$.
The boundary is then defined to be
zero level set of this new function, which automatically
satisfies Dirichlet boundary conditions, and the
domain will be the interior of this boundary.

In principle any choice of flow and an appropriate
choice of harmonic function may work.
For simplicity we begin with the ellipse flow
and modify it with a harmonic polynomial.
Up to overall multiplicative constants, the
new (non-dimensional) flow solution can be
written as
\begin{equation}
u(y,z) = 1 - c_1 \left( y^2 + c_2^2 z^2 \right) - c_3 P(y,z)
\end{equation}
with harmonic polynomial $P(y,z)$ and undetermined
coefficients $c_1$, $c_2$, and $c_3$.
This is a generalization of a few different
geometries. For example, the
ellipses can be obtained by setting
$(c_1,c_2,c_3) = (1,\lambda,0)$, and
the infinite channel can be obtained with
$(c_1,c_2,c_3) = (1,0,0)$.

\begin{figure}
\centering
\includegraphics[width=\linewidth]{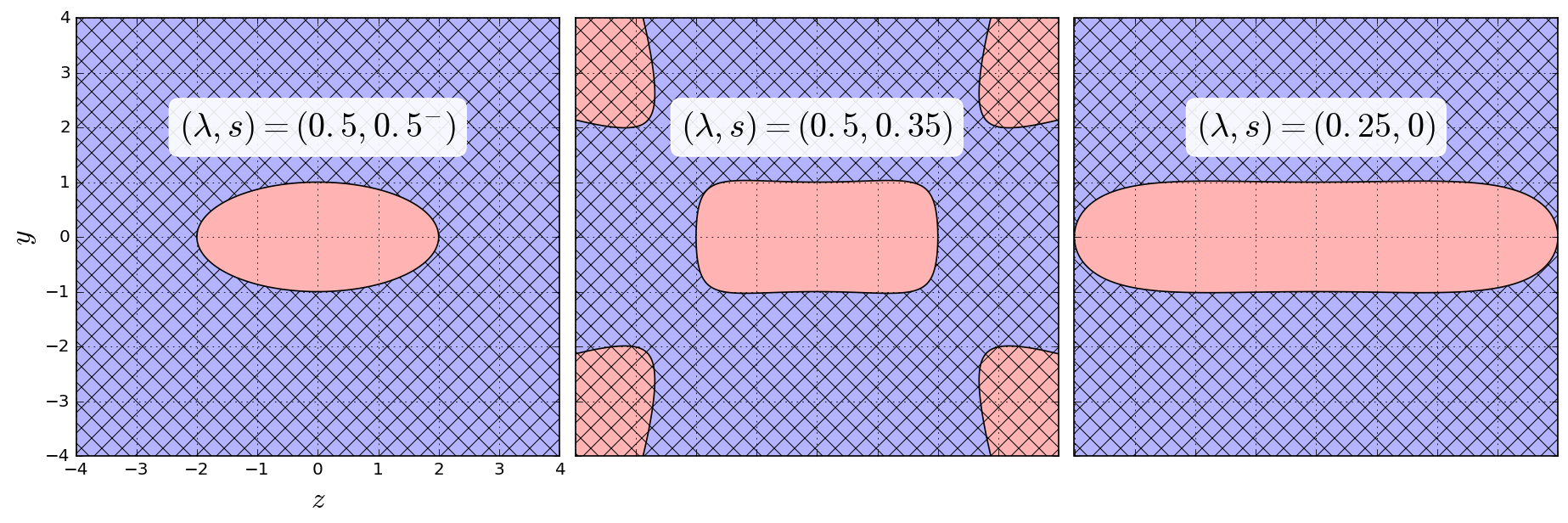}
\caption{
Illustration of the racetrack for the choices
of parameter pairs shown. Note that for $s<\lambda$
it is common for the domain to be come
non-convex, as can be seen with $(\lambda,s) = (0.5,0.35)$.
The hatched regions indicate the exterior of the domain,
and red and blue indicate regions where $u>0$ and
$u<0$ respectively.
}
\label{fig:rt_examples}
\end{figure}

Now we seek an appropriate polynomial.
Due to the Cauchy-Riemann equations,
the only polynomials of two variables
which are harmonic are linear combinations of
the real and imaginary parts
of complex polynomials $f(w) = w^n = (y + i z)^n$.
The simplest non-trivial polynomial of this class with
four-fold symmetry is
\begin{equation}
P(y,z) = \Re \left[ (y+i z)^4 \right] = y^4 - 6y^2 z^2 + z^4.
\end{equation}
Using this $P(y,z)$, we are left with specifying
the undetermined constants to ensure the domain
has the right properties.
We impose the conditions $u(1,0) = u(0,1/\lambda) = 0$
to maintain the aspect ratio.
%
\begin{equation}
\left.
\begin{array}{r l}
u(1,0) &= 1-c_1 - c_3 = 0 \\
u(0,1/\lambda) &= 1-\frac{c_1 c_2^2}{\lambda^2} - \frac{c_3}{\lambda^4} = 0
\end{array}
\right\} \Rightarrow
\left\{
\begin{array}{r l}
c_1 &= \frac{1-\lambda^4}{1-\lambda^2 c_2^2} \\
c_3 &= 1-c_1 = \frac{\lambda^2(\lambda^2 - c_2^2)}{1-\lambda^2 c_2^2}
\end{array}
\right.
\end{equation}
With two equations and three coefficients,
we redefine $c_2 \equiv s$ to use as a shape parameter.
Substituting, the resulting flow with shape parameter $s$ is
\begin{equation}
\label{eqn:u_racetrack}
u(y,z;s) = 1 - \left(\frac{1-\lambda^4}{1-\lambda^2 s^2}\right) \left( y^2 + s^2 z^2 \right) - \left(\frac{\lambda^2(\lambda^2 - s^2)}{1-\lambda^2 s^2}\right) (y^4 -6y^2 z^2 + z^4),
\end{equation}
and by construction, this choice of $u$ satisfies the Poisson problem
\begin{equation}
\begin{aligned}
\label{eqn:racetrack_problem}
&\Delta u = const., \qquad u|_{\partial \Omega(s)} = 0, \\
&\Omega(s) \equiv \left\{(y,z) \, : \, u(y,z;s) \geq 0 \right \}.
\end{aligned}
\end{equation}
The domains for a few choices of parameter
pairs $(\lambda,s)$ are shown in figure \ref{fig:rt_examples}.
Note, the diagonal, $s=\lambda$, parametrizes pure ellipses of different aspect ratio, limiting to a circle at $s=\lambda=1$.
Due to the nature of the harmonic
perturbation, there are disconnected regions
for which $u>0$, so some care must be taken.
If the harmonic component (whose level sets are unbounded) is
large relative to the elliptical component, we expect the
level sets of their sum to be unbounded.
For example, this happens in the regime $\lambda \approx 1$
and $s \ll \lambda$.
We have taken care to ensure
our results are only in physically-relevant domains.
For extreme (yet still physical) values of $s$, the
domain will become non-convex, the four ``probes"
on the exterior, as seen in the center panel,
come towards the corners of
the domain, which results in thin regions of high curvature,
which makes both aspects of the numerics more challenging.

\subsection{Monte Carlo simulation for the racetrack}

We use the same Monte Carlo technique described in \cite{Science2016},
where the full tracer solution is
the probability density corresponding to sample
paths of the stochastic differential equation
\begin{equation}
d\mathbf{X}(t) = \text{Pe} \, \mathbf{u}(Y(t),Z(t))dt + d\mathbf{W}(t).
\end{equation}
The racetrack flow is a polynomial by construction,
so it can be evaluated directly.
The boundary is defined implicitly by $u(y,z)=0$, and, except for
a measure-zero set of parameters ($\lambda$,s), any physically
relevant domain has a region
for which $u<0$ in the exterior surrounding the domain
(see figure \ref{fig:rt_examples}).
This gives us a way to detect when a particle $\mathbf{X}$ has exited
the domain, and allows
us to use a similar algorithm to perform
reflections off the boundary using $\underline{n} \propto -\nabla u$.
The main complication when working with an implicit domain
is finding the solution(s) of $u(f(s),g(s))=0$, the location
of the crossing, which requires a
numerical rootfinding step for these quartic polynomials along
the parameterized line segment $(f(s),g(s)), \; \; s \in [0,1]$
(not to be confused with shape parameter $s$).

While the flow formula is well behaved for any $(\lambda,s)$
(except for $\lambda=s=1$ as a special case),
the domain shape varies drastically depending on the
pair chosen. First, one must restrict to $s \leq \lambda$,
otherwise the domain will not be as expected,
as the ``pinning" of the domain to $(y,z) = (1,0)$ and
$(0,1/\lambda)$ will be satisfied, but occur on
algebraic curves outside the domain, instead of the main racetrack.
Secondly, for sufficiently small $s$ and moderate $\lambda$,
the domain can become unbounded. For moderately small $s$,
the special care with the reflection
algorithm must be taken to ensure the physics is properly resolved.

\begin{figure}
\centering
\includegraphics[width=\linewidth]{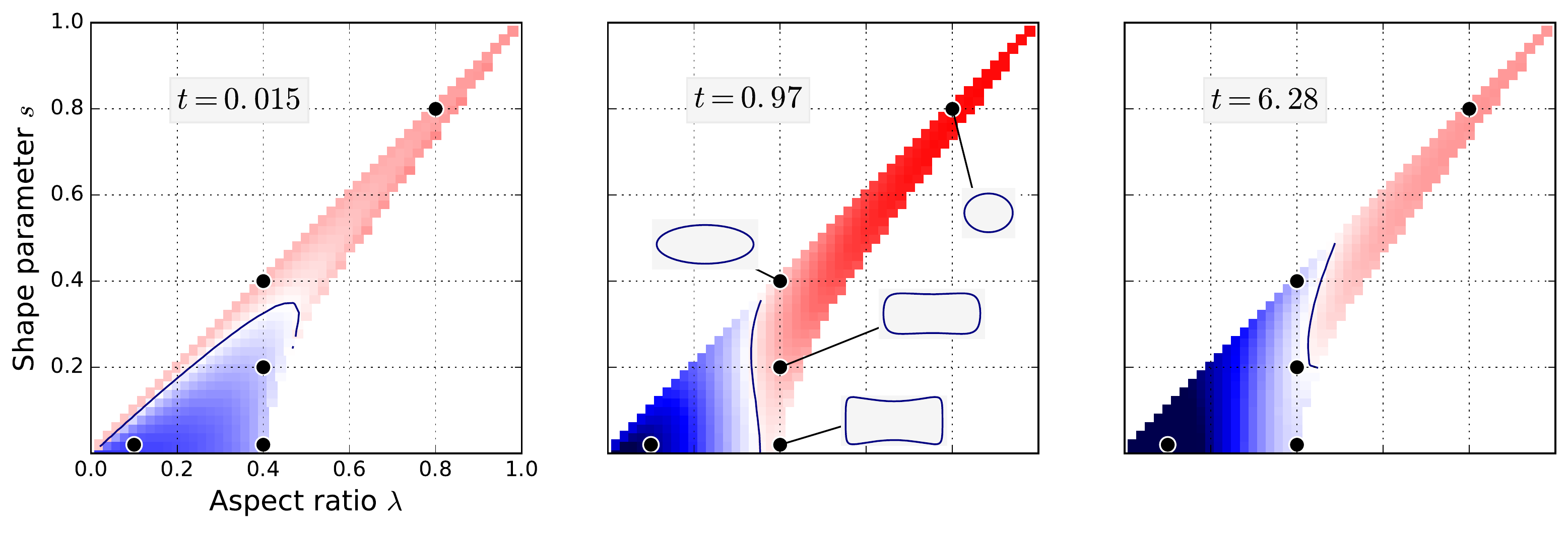}
\begin{tabular}{| c | c | c | c | c | c | c |}
\hline
$(\lambda,s)$ & $(0.1,0.02)$ & $(0.4,0.02)$ & $(0.4,0.2)$ & $(0.4,0.4)$ & $(0.8,0.8)$  \\
\hline
$\mathcal{S}^\mathcal{G}$ & $-0.2427$ & $-0.1020$ & $-0.1171$ & 0 & 0  \\
\hline
LT coeff & $-4.5415$ & $-0.1150$ & $-0.0587$ & $-0.2131$ & $0.2446$  \\
\hline
\end{tabular}
\caption{Top Row:
Averaged skewness over a range of parameter pairs
$(\lambda,s)$ which satisfy a boundedness of domain criterion, plotted at nondimensional times
(from left to right)
$t \approx 0.015$,  $0.97$, and $6.28$.
Positive (negative) skewness is red (blue),
with white being zero. An approximate zero skewness
contour is overlaid in black. Long time behavior
is seen to be nearly independent of the
shape parameter. Bottom Row:  Table of short (exact) and long time (with finite elements) skewness values at the labeled points in the left panel of the top row. Sample shapes are reported for located for the marked spots in the $(\lambda,s)$ plane.
}
\label{fig:rt_paramsweep}
\end{figure}

A summarized result of a parameter sweep in $(\lambda,s)$
space for the racetracks is shown in figure \ref{fig:rt_paramsweep}.
Nonphysical parameter pairs have been thrown out, as
judged by anomalously large negative skewness.
The panels correspond to three snapshots in time, at
$t \approx 0.015$,  $0.97$, and $6.28$. The line $s=\lambda$
corresponds to the ellipses; vertical lines $\lambda=\textit{const.}$
explore various shapes of a fixed aspect ratio.
A uniform colormap is used across all images,
where blue, red, and white correspond
to negative, positive, and zero skewness, respectively,
with the approximate zero-skewness curve
drawn in black.
At $t=0.015$, the after-effects of geometric
skewness are seen; the ellipses ($\lambda=s$)
are all positive, while the rectangular-like shapes
(bottom boundary) are separated between
positive and negative values between $\lambda=0.4$ and
$\lambda=0.5$. Approaching the
diffusive timescales in the center and right panels,
the behavior becomes nearly independent of the shape
parameter, providing evidence that the ``golden" long
time aspect ratio of $\lambda^{*} \approx 0.49$ is observed across
this class of geometries.

The table in figure \ref{fig:rt_paramsweep} shows the
numerically computed asymptotic coefficients for a
few choices of parameters, marked with dots in the figure.
The coefficients are found using the finite element
package available in Mathematica 10,
and the solver parameters are the same as those
used in the polygons discussed in the previous section.
In contrast to the polygons, the
meshes here are formed using built-in tools to define
implicitly defined domain; in this case, $u \geq 0$.

These results suggest that
in applications which reach at least the first diffusive timescale,
all else being equal, the behavior of asymmetry is
insensitive to the details of the cross-sectional shape.

\section{Skewness as a measure of mass asymmetry}

In elementary statistics, a nonzero centered skewness is often taken  to be
the measure of a distribution's asymmetry. While
it is not too difficult to construct counterexamples
to this intuition, namely non-symmetric distributions whose skewness is
zero, with relatively tame additional assumptions
on the form of a distribution it can be proved
that the skewness, and the
``loading,", i.e., the median, of the distribution
have opposite sign, thus confirming the intuition for a
large class of statistics. In this section, we
give a proof and isolate a sufficient requirement, which we will
refer to as ``single-crossing," for the statistical skewness
to agree with such intuition.
We also monitor the sign of the median
and the sign of the skewness in the case
of the triangle, and will observe timescales on which the single crossing condition does not hold. Finally, we give domain-independent
arguments to show that at both sufficiently large
and sufficiently small times cross-sectionally
averaged tracer distributions will always satisfy
the single-crossing condition.

\subsection{Single crossing distributions}
\begin{figure}
\centering
\includegraphics[width=3in]{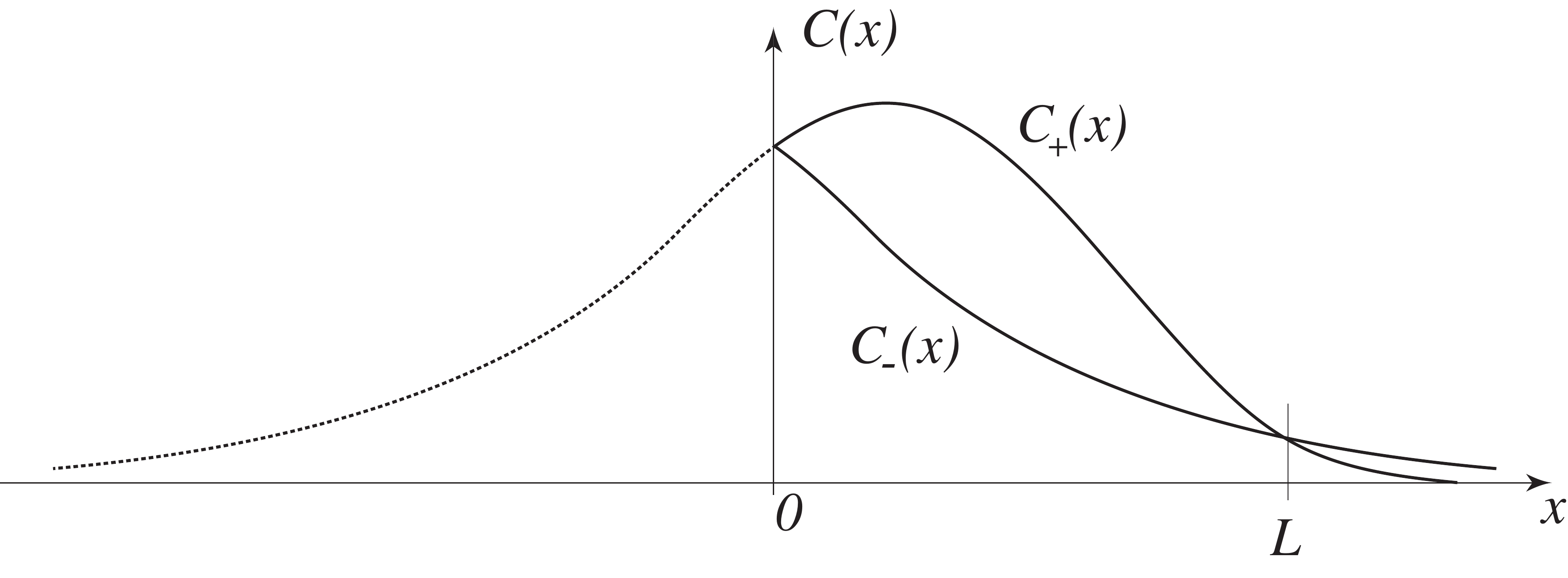}
\caption{
Single crossing for reflected distribution.
}
\label{snglcrss}
\end{figure}

Let a unimodal distribution $C(x)$, supported on the real line $x \in \barr$,  be such that the origin $x=0$ coincides with the mean (center of mass) of the distribution, i.e.,
$$
\left. X\equiv \int_{-\infty}^{+\infty} x C(x) \, d x \right/ \int_{-\infty}^{+\infty} C(x) \, d x  =0 \, ,
$$
so that
$$
\int^{+\infty}_{0} x\, C(x) \, d x =-\int^0_{-\infty}x\, C(x) \, d x\, ,
$$
hence
$$
\int_0^{+\infty}x\, C(x) \, d x=\int_0^{+\infty}x\, C(-x) d x \,.
$$
Define
$$
C_+(x)\equiv C(x) \quad {\rm and} \quad C_-(x)\equiv C(-x)\, ,  \qquad 0<x<\infty \, ,
$$
and assume that the graphs of $C_+$ and $C_-$ cross transversally at a single point $x=L$, with their relative magnitudes as exemplified
by figure~\ref{snglcrss}. Thus, $C_-$ is smaller than the peak of the distribution attained by $C_+$ in this example, and zero mean is achieved with a longer decaying tail for $C_-$ as $x$ increases towards infinity (which implies at least one crossing of the two graphs).
Then, upon rescaling $\tilde{x}\equiv x/L$, we have
$$
\int_0^{1}\tilde{x}\, C_+(\tilde{x}) \, d \tilde{x}+\int_1^{\infty}\tilde{x}\, C_+(\tilde{x}) \, d \tilde{x}=
\int_0^{1}\tilde{x}\, C_-(\tilde{x}) \, d \tilde{x}+\int_1^{\infty}\tilde{x}\, C_-(\tilde{x}) \, d \tilde{x}
$$
or
$$
\int_0^{1}\tilde{x}\, \Big(C_+(\tilde{x}) -C_-(\tilde{x})\Big) \, d \tilde{x}=
\int_1^{\infty}\tilde{x}\, \Big(C_-(\tilde{x}) -C_+(\tilde{x}) \Big)\, d \tilde{x}
$$
and the inequalities
$$
C_+(\tilde{x})>C_-(\tilde{x})\, \quad 0<\tilde{x}<1
\qquad {\rm and} \qquad
C_+(\tilde{x})<C_-(\tilde{x})\,, \quad 1<\tilde{x}<\infty\, ,
$$
guarantee that the integrands on both sides of the last equality are non-negative.
Hence, the following inequalities hold,
\begin{eqnarray}
&&\hspace{-2.5cm}
\int_0^{1}\Big(C_+(\tilde{x}) -C_-(\tilde{x})\Big) \, d \tilde{x}>
\int_0^{1}\tilde{x}\Big(C_+(\tilde{x}) -C_-(\tilde{x})\Big) \, d \tilde{x}=
\nonumber
\\
&&=\int_1^{\infty}\tilde{x}\Big(C_-(\tilde{x}) -C_+(\tilde{x}) \Big)\, d \tilde{x}
>\int_1^{\infty}\Big(C_-(\tilde{x}) -C_+(\tilde{x}) \Big)\, d \tilde{x}\, ,
\end{eqnarray}
which, upon rearranging terms, rescaling back to the original variables and using the definition of $C_\pm$, imply
$$
\int_0^{+\infty}  C(x) \, d x >  \int_{-\infty}^{0}   C(x) \, d x \, ,
$$
that is, the mass of the distribution to the right of the origin (by assumption chosen to be the center of mass, or mean, of the distribution)
is larger than the mass to the left of the origin. Thus,  the median under the single crossing requirement and the relative magnitudes illustrated by figure~\ref{snglcrss} lies to the right of the mean,  that is, with respect to this position the distribution is ``front loaded."

Similarly, working with  the third moment  leads to the chain of inequalities
\begin{eqnarray}
&&\hspace{-2.5cm}
\int_0^{1}\tilde{x}^3\Big(C_+(\tilde{x}) -C_-(\tilde{x})\Big) \, d \tilde{x}<
\int_0^{1}\tilde{x}\, \Big(C_+(\tilde{x}) -C_-(\tilde{x})\Big) \, d \tilde{x}=
\nonumber
\\
&&=\int_1^{\infty}\tilde{x}\, \Big(C_-(\tilde{x}) -C_+(\tilde{x}) \Big)\, d \tilde{x}
<\int_1^{\infty}\tilde{x}^3\Big(C_-(\tilde{x}) -C_+(\tilde{x}) \Big)\, d \tilde{x}\, ,
\end{eqnarray}
and rearranging terms again shows that the third moment is negative with respect to the origin set by the mean,
$$
\int_{-\infty}^{\infty}x^3 C(x)  d x<0 \, .
$$
Hence,  negative skewness implies front loaded distributions under the unimodal requirement and single transverse crossing, with respect to the origin set by the mean, of their reflected left side with the right side.

Armed with this mathematical criterion, it is an interesting question to consider when it may be satisfied for the passive scalar evolution.  To this end, define the symmetric and anti-symmetric modes, $T_+$, $T_-$:
\begin{eqnarray}
T_+&=& T(x,y,t) + T(-x,y,t)\\
T_-&=& T(x,y,t) - T(-x,y,t)
\end{eqnarray}
These each satisfy the coupled system:
\begin{eqnarray}
\frac{\partial T_+}{\partial t} - \Delta T_+&=& -\Pen u(y,z) \frac{\partial T_-}{\partial x}\\
\frac{\partial T_-}{\partial t} - \Delta T_-&=& -\Pen u(y,z) \frac{\partial T_+}{\partial x}
\end{eqnarray}
Performing Picard iteration, say assuming a small $\Pen$ expansion, will yield short time asymptotics:
\begin{eqnarray}
T_+&=& {T_+}^{(0)} + \Pen {T_+}^{(1)} + \cdots\\
T_-&=& {T_-}^{(0)} + \Pen {T_-}^{(1)} + \cdots\\
\end{eqnarray}
Note, that for the case of symmetric initial data with respect to $x$, the first term,
${T_-}^{(1)}$ is zero, and the iteration proceeds as:
\begin{eqnarray}
\frac{\partial {T_+}^{(0)}}{\partial t} - \Delta {T_+}^{(0)}&=& 0\\
\frac{\partial {T_-}^{(1)}}{\partial t} - \Delta {T_-}^{(1)}&=&- u(y,z) \frac{\partial {T_+}^{(0)}}{\partial x}\\
\frac{\partial {T_+}^{(1)}}{\partial t} - \Delta {T_+}^{(1)}&=&- u(y,z) \frac{\partial {T_-}^{(0)}}{\partial x}
\end{eqnarray}
Separating the $x$ variables from the others, assuming the initial data is a function only of $x$, the leading order symmetric mode is $2 G(x,t)$, where $G(x,t)$ is the solution of the one space, one time free space Green's function for the heat equation.
Continuing with the variable separation, ${T_-}^{(1)}=2 H(y,z,t) G_x(x,t)$, produces
\begin{eqnarray}
\frac{\partial H}{\partial t} - (H_{yy}+H_{zz}) &=&- u(y,z)
\end{eqnarray}
The short asymptotic solution for this problem is $H(y,z,t)\sim -t u(y,z)$ in the bulk.
Note that there is a missing boundary layer from the vanishing no-flux condition, but this does not contribute to the cross sectionally averaged skewness at leading order in short time.

Repeating the procedure and inverting, yields:
\begin{eqnarray}
T(x,y,z,t)\sim G-\Pen t u G_x+ \frac{1}{2}\Pen^2t^2\left(u^2 G_{xx}-\frac{1}{3}\Pen t u^3 G_{xxx}\right)\\
T(-x,y,z,t)\sim G+\Pen t u G_x+\frac{1}{2}\Pen^2t^2\left(u^2 G_{xx}+\frac{1}{3}\Pen t u^3 G_{xxx}\right)
\end{eqnarray}
Since the cross sectional average of $u$ is zero, and assuming that the cross sectional average of $u^3$ is non-zero, determining the number of zero crossings for positive $x$ corresponds to finding the zeros of the function $G_{xxx}$.  For Gaussian initial data, this point is unique, and corresponds to the inflection point of $G_x$.  This establishes that the first symmetry breaking at short time involves a single crossing for our passive scalar equation, and hence the skewness sign indicates the front versus back loading properties of the distribution.  Similar arguments can be provided for the long time counterpart using methods of homogenization which we will report in future work.

\begin{figure}
\centering
\includegraphics[width=5in]{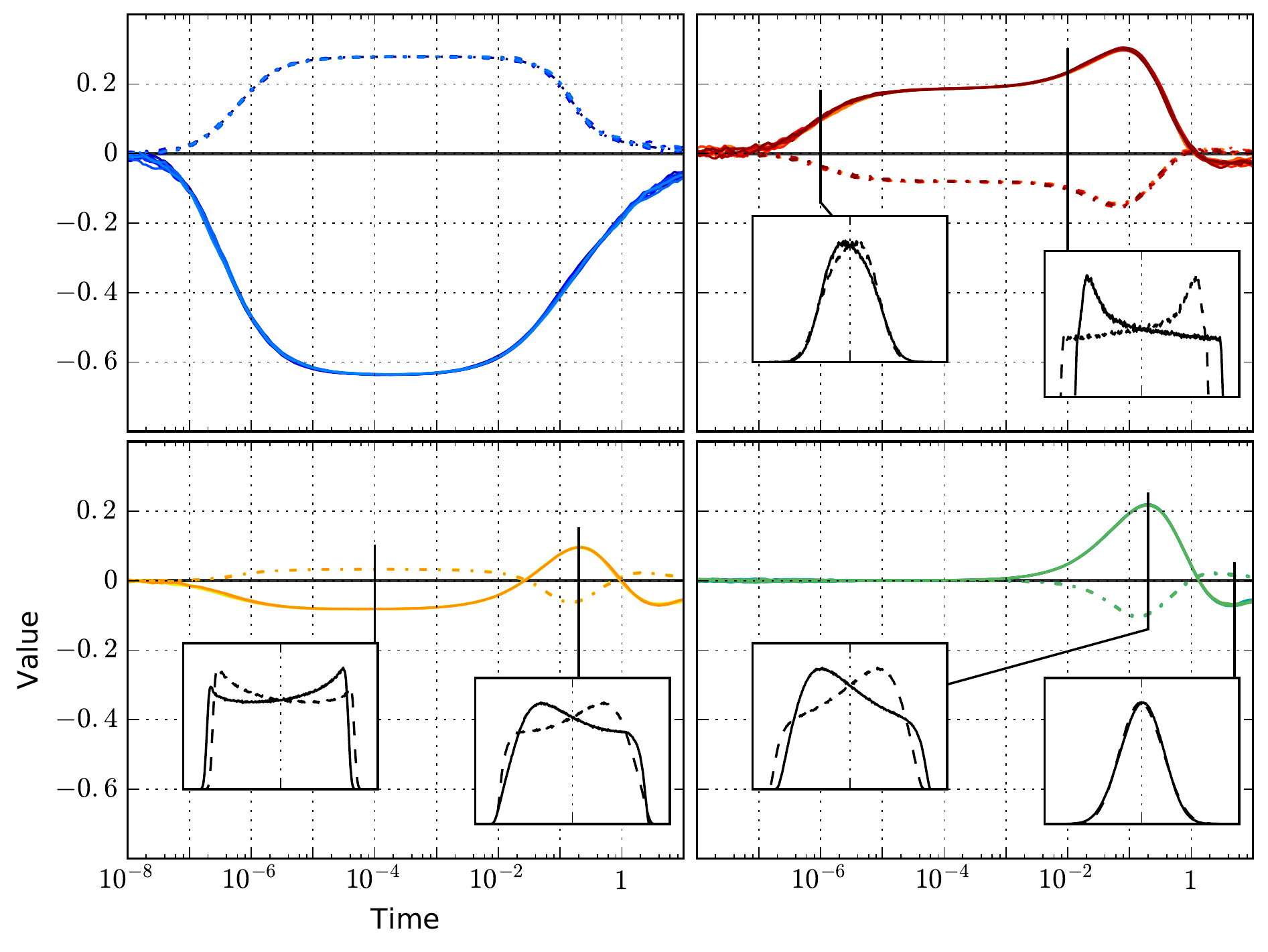}
\caption{Evolution of Median and Skewness: top left is channel, bottom left is duct and bottom right is ellipse each of aspect ratio $0.4$, while top right is triangle.  Dotted curves are median scaled by standard deviation, solid curves are skewness.  Insets show the cross-sectionally average distributions (solid), and its reflection about the mean (dots) as a function of longitudinal direction.  Observe that there are time intervals of multiple crossings, while the anti-correlation between median and skewness prevails except for a few extremely short time intervals for the right panels at later times.
}
\label{median}
\end{figure}

In the left panel of figure \ref{median}, we plot the evolution of the skewness and the median of the distribution for the 4 cases:  the triangle, an ellipse of aspect ratio $0.4$, a rectangular duct of same aspect ratio, and the case of the infinite channel.  Observe the strong opposite sign correlation between the median and skewness for essentially all times documented.  We remark that there are times for which the cross-sectionally average distribution is observed in the Monte-Carlo simulations to have multiple crossings with its reflected graph, as depicted in the insets of this figure.  Nonetheless, the strong anti-correlation persists, illustrating how the single crossing conditions presented here is sufficient but not necessary.  The discerning reader will observe that there is an extremely short time interval in the case of the triangle (upper right), and elliptical domain (lower right) near the diffusion timescale for which the anti-correlation between median and skewness appears to be violated.  Examining the intersection count in this case is inconclusive as the distributions are extremely near Gaussian in this time interval, and much higher fidelity simulations are needed to resolve this behavior.

\section{Conclusion}

We have demonstrated that for long times, the triangle stands apart from all other regular polygons in that the concentration distribution's skewness approaches zero through negative values, despite its short time sign behavior agreeing with all other regular polygons.
We have also explored smooth deformations of elliptical domains which mimic rounded rectangles and exact flow solutions which demonstrate that the skewness signature does not necessarily correlate with the presence of corners in the cross-sectional domain, but is set primarily by the aspect ratio.  Lastly, we explore a criterion which guarantees when the skewness sign indicates a mass loading property of the concentration.  We establish that this condition is satisfied quite generically at short (and long) times.  All these observations are supplemented by direct Monte-Carlo simulations which document applicability of these results in finite time as needed in real experiments.

\section{Acknowledgements}

We thank Ian Griffiths for suggesting we extend our studies to the triangle, and pointing out that an explicit polynomial solution is available for the flow in that case.
We acknowledge funding from the Office of Naval Research (grant
DURIP N00014-12-1-0749) and the NSF (grants RTG
DMS-0943851, CMG ARC-1025523, DMS-1009750, and
DMS-1517879).

\bibliographystyle{plain}

\end{document}